\documentclass[prl,twocolumn,superscriptaddress,preprintnumbers,amsmath,amssymb,floatfix]{revtex4}
\usepackage{graphicx}

\begin{document}
\title{Non-quasiparticle states in Co$_2$MnSi evidenced through 
magnetic tunnel junction spectroscopy measurements}
\author{L. Chioncel}
\affiliation{Institute of Theoretical Physics, Graz University of Technology,
A-8010 Graz, Austria}
\affiliation{Faculty of Science, University of Oradea, RO-410087 Oradea, Romania}
\author{Y. Sakuraba}
\affiliation{Department of Applied Physics, Graduate School of
Engineering, Tohoku University, Sendai 980-8579, Japan}
\author{E. Arrigoni}
\affiliation{Institute of Theoretical Physics, Graz University of Technology,
A-8010 Graz, Austria}
\author{M.I.~Katsnelson}
\affiliation{Institute for Molecules and Materials, Radboud
University of Nijmegen, NL-6525 ED Nijmegen, The Netherlands}
\author{M.~Oogane}
\affiliation{Department of Applied Physics, Graduate School of
Engineering, Tohoku University, Sendai 980-8579, Japan}
\author{Y. Ando}
\affiliation{Department of Applied Physics, Graduate School of
Engineering, Tohoku University, Sendai 980-8579, Japan}
\author{T. Miyazaki}
\affiliation{Department of Applied Physics, Graduate School of
Engineering, Tohoku University, Sendai 980-8579, Japan}
\author{E. Burzo}
\affiliation{Babe\c s-Bolyai University, RO-800084 Cluj-Napoca, Romania}
\author{A.I.~Lichtenstein}
\affiliation{Institute of Theoretical Physics, University of Hamburg,  DE-20355 Hamburg, Germany}

\begin{abstract}
We investigate the effects of electronic correlations in the full-Heusler 
Co$_2$MnSi, by combining a theoretical analysis of the spin-resolved density 
of states with tunneling-conductance spectroscopy measurements  using Co$_2$MnSi 
as electrode. Both experimental and theoretical results confirm the existence 
of so-called non-quasiparticle states and their crucial contribution 
to the finite-temperature spin polarisation in this material. 
\end{abstract}


\maketitle

Next generation electronic devices will profit from technologies
that control the spin degrees of freedom. Therefore, half-metallic
ferromagnets (HMF), in which the majority-spin density of states (DOS)
crosses the Fermi level ($E_F$) while the minority-spin DOS shows a
semiconducting behavior at $E_F$ (or vice versa), can be seen as
essential components for tunneling magnetoresistance (TMR)
devices. In HMF-based TMR devices the magnetoresistance 
should, ideally, diverge if the conduction electron spins are
$100\%$ spin-polarized.  It is further expected that, if such a HMF
could be epitaxially grown on a semiconducting surface, fully
polarized ($100\%$) electrons could be possibly injected into the
semiconductor.

First band-structure calculations performed on the half-Heusler, NiMnSb 
(C1$_b$-type structure), predicted a $100\%$ spin polarization \cite{gr.mu.83}.
Subsequently, the studies were extended to other half-metallic systems. 
Of particular interest for realizing magnetic tunneling junctions (MTJ)
appears to be the full-Heusler alloy Co$_2$MnSi (L2$_1$-type structure). A 
minority-spin band-gap of 0.4eV has been predicted and a Curie temperature
of $T_c=985K$ and a saturation magnetisation of $5\mu_B$ was reported 
\cite{ra.ra.01}. 

Some of us \cite{sa.mi.06} recently fabricated magnetic
tunnel junctions (MTJ) consisting of highly ordered Co$_2$MnSi
epitaxial bottom electrode, Al-O tunnel barrier, and
Co$_{75}$Fe$_{25}$ top electrode. A TMR ratio of
$159\%$, at low temperatures and a value of $70\%$ at room
temperature, was determined. More recently, MTJ
structures consisting of Co$_2$MnSi/Al-O/Co$_2$MnSi were
fabricated, having a TMR ratio of $570\%$, at 2K, the 
largest one reported to date for an Al-O amorphous
tunneling barrier \cite{sa.ha.06}. These experiments reveal the
HMF character of Co$_2$MnSi with a minority spin band gap 
and  a high decrease of 
TMR ratio with temperature \cite{sa.ha.07}.

In order to understand the large temperature variation of the TMR
ratio in Co$_2$MnSi, 
it is important to investigate the temperature dependence
of its half-metallic density of states
in the presence of electron-electron interaction. 
Zero-temperature band structure calculations 
within the framework of Density Functional Theory (DFT) were reported 
by
Galanakis et al. \cite{ga.ma.06}. According to these calculations, the
half-metallic character 
of Co$_2$MnSi
is determined by the existence of a minority 
spin gap formed between the triply degenerate Co-Co anti-bonding t$_{2g}$ 
and the double degenerate Co-Co anti-bonding e$_g$ \cite{ga.ma.06} bands. 
However, standard DFT  - local density approximation (LDA), calculations are 
in general insufficient to describe some important many-body features of 
HMF, at zero or finite temperatures. One of these effects is the appearence 
of so-called non-quasiparticle (NQP) states, 
i. e. states appearing within the minority spin band 
gap just above the Fermi level. 
These states describe
low-energy electron excitations for minority spins, 
which
turn out to be possible as 
superpositions of spin-up electron excitations and virtual magnons 
\cite{ed.he.73,ir.ka.90,ir.ka.94} (``spin-polaron'') processes.
Therefore, their description require an appropriate treatment of
dynamical many-body effects.
NQP states 
have been studied in several half-metals 
\cite{ch.ka.03,ch.ar.06,ch.ka.05,ch.ma.06,ch.al.07}
by using a combined LDA+DMFT 
(dynamical mean field theory) approach (for a review, see
Ref.\onlinecite{ko.sa.06}).
NQP states contribute significantly to the tunneling transport in  heterostructures
based on HMF \cite{ir.ka.02,mc.fa.02,ir.ka.06}.
At $T=0K$, the density of NQP  states vanishes at the Fermi level,
$E_F$ 
(in the presence of spin anisotropy
at a slightly higher energy $E_F+\hbar
\omega_m$, see below), 
while for $T>0$ tails of the NQP states cross the Fermi
energy and contribute to the depolarization.

In this Letter, we show the crucial importance of non-quasiparticle 
states for the finite-temperature depolarisation in Co$_2$MnSi.
This is achieved by combining a theoretical analysis of the spin-resolved density 
of states with tunneling-conductance spectroscopy measurements at different 
temperatures in a Co$_2$MnSi - based MTJ. 
Our combined experimental and theoretical analysis confirms the 
presence of NQP states and emphasizes their contribution to 
the finite temperature polarization in this material.


Band-structure calculations were performed for the half-metallic Co$_2$MnSi 
alloy
for the experimental lattice constant $a=5.65\AA$. 
Calculations were performed using a recently developed LSDA+DMFT scheme 
\cite{ch.vi.03}. 
Correlation effects in the valence Co 
and Mn $d$ orbitals are included via an on-site electron-electron interaction 
in the form
$\frac{1}{2}\sum_{{i \{m, \sigma \} }} U_{mm'm''m'''}
 c^{\dag}_{im\sigma}c^{\dag}_{im'\sigma'}c_{im'''\sigma'}c_{im''\sigma} $.
The interaction is treated in the framework of dynamical mean field theory (DMFT)
\cite{ko.sa.06}, with a spin-polarized T-matrix Fluctuation Exchange (SPTF) type 
of impurits solver \cite{ka.li.02}. 
Here, $c_{im\sigma}/c^\dagger_{im\sigma}$ 
destroys/creates an electron with spin $\sigma$ on orbital $m$ on site $i$.
The Coulomb matrix elements $U_{mm'm''m'''}$ are expressed in the usual 
way~\cite{im.fu.98} 
in terms of  three Kanamori parameters 
$U$, $U'$ 
and $J$.
Since
the static part of 
the correlation effects is already included in the local spin-density 
approximation (LSDA), ``double counted'' terms must be subtracted. To obtain this, 
we replace $\Sigma_{\sigma}(E)$ with $\Sigma_{\sigma}(E)-\Sigma_{\sigma}(0)$
\cite{li.ka.01}  
in all equations of the LSDA+DMFT procedure \cite{ko.sa.06}. 
Physically, this is related to the fact that DMFT only adds {\it dynamical}
correlations to the LSDA result. 
For this reason, it is believed that this kind 
of double-counting subtraction ``$\Sigma(0)$'' is more appropriate for a DMFT 
treatment of metals than the alternative static ``Hartree-Fock'' (HF) subtraction
\cite{pe.ma.03}.

\begin{figure}[h]
\includegraphics[width=0.95\linewidth, height=7cm]{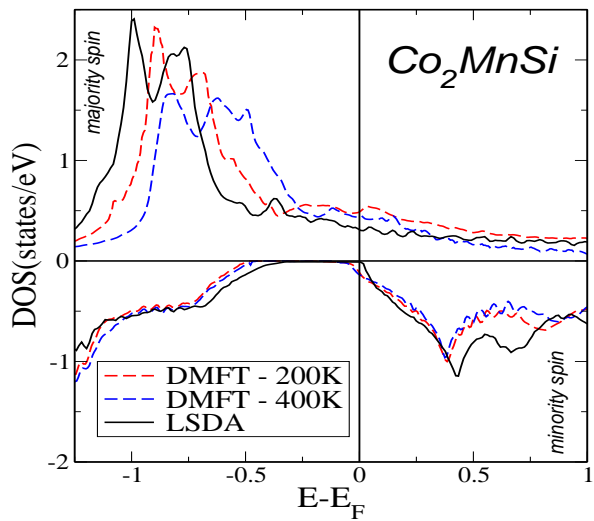}
\caption{(Color online) Spin resolved density of states of the Co$_2$MnSi 
full-Heusler alloy obtained by LSDA and LSDA+DMFT. The LSDA+DMFT results 
are obtained for U=3eV and J=0.9eV (in both Co and Mn), and different temperatures.}  
\label{dos-Co2MnSi}
\end{figure}
Fig. \ref{dos-Co2MnSi} shows the results of DOS calculations using the LSDA 
and LSDA+DMFT schemes at different temperatures. Realistic values  of $U$ 
for all 3d metals are predicted to vary between 2 and 6-7 eV \cite{ko.sa.06}. 
We have checked that for values of the $U$ parameters between 2 and 4eV  the 
spectral weight of NQP states is not significantly changed in agreement with 
recent calculations \cite{ch.ar.06}. Here, we present results for 
$U=3eV$ and $J=0.9eV$. In order to resolve the LSDA minority-spin gap, a   
broadening of around 15K and a k-point mesh of 2304 points was used in the 
reduced Brillouin zone.
The LSDA density of states, i. e. the $T=0K$ result, confirms the existence of a
minority-spin gap at $E_F$, in agreement with previous results \cite{ga.ma.06}. In 
contrast, finite-temperature results obtained within LSDA+DMFT display a 
broadening of the NQP states across the Fermi energy (Fig. \ref{dos-Co2MnSi}) 
in the minority spin channel, and a spectral weight redistribution for
majority spins. 

Static non-collinear spin configurations  at finite temperatures, produce a 
homogeneous mixture of spin-up and spin-down density of states, in such a way that the 
proportionality relation between polarization and magnetization is roughly 
maintained as a function of temperature $\delta P(T) \propto \delta<S^z>$ \cite{sk.do.02}.
Previously, this proportionality relation was reproduced in  magnetic semiconductors
\cite{edwa.83}, in qualitative agreement with experimental data \cite{ki.ba.78}. In 
contrast, for some HMF materials, it was shown by model considerations \cite{ir.ka.06} 
as well as direct numerical calculations \cite{ch.ar.06,ch.ka.03,ch.al.07}, 
that the polarization displays a completely different temperature behavior in comparison 
with the magnetization. 


\begin{figure}[h]
\includegraphics[width=0.85\linewidth, height=9.5cm]{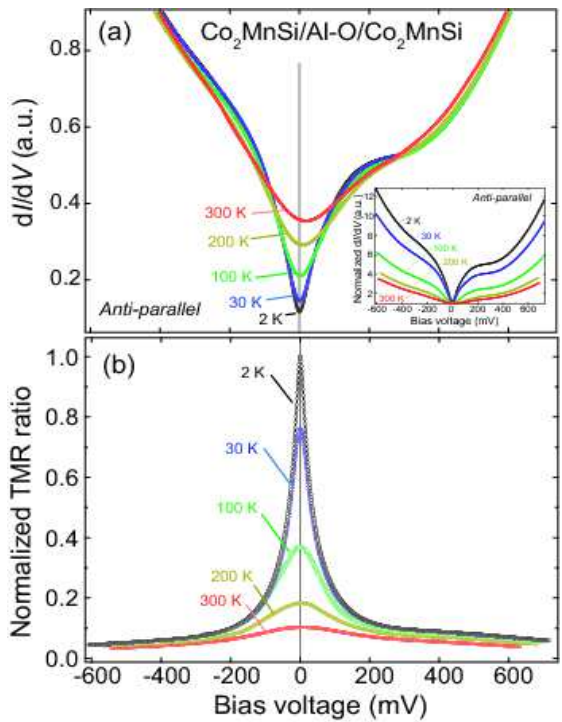}
\caption{(Color online) (a) Tunnel conductance for the
Co$_2$MnSi/Al-O/Co$_2$MnSi anti-parallel magnetic tunnel junction
and (b) 
TMR ratio (normalized to $T=2K, V=0$), as a function of voltage for different
temperatures. 
The inset of
panel (a) shows the conductance normalized to zero bias voltage.}
\label{mtj-Co2MnSi}
\end{figure}

In order to confirm experimentally the existence of NQP states in Co$_2$MnSi and to 
investigate their  temperature dependence, we carried out tunneling spectroscopy 
measurements for Co$_2$MnSi-based MTJ. 
Specifically, the measurements
were 
first
carried out for Co$_2$MnSi/Al-O/Co$_2$MnSi-MTJ, for which
a TMR ratio of 570$\%$ at 2K was achieved in previous
studies~\cite{sa.ha.07}.
A positive bias voltage was applied to the upper Co$_2$MnSi electrode,
in order to control the tunneling of electrons from the lower electrode to the 
upper one. 

Figs. \ref{mtj-Co2MnSi} (a) and (b) show the bias-voltage dependence of the 
differential tunneling conductance 
($dI/dV$) 
in the anti-parallel (AP) 
configuration, and the normalized TMR ratio, respectively. The measurements were 
conducted at temperatures between $2 K$ and $300 K$. 
At $T =2K$, only a very small 
tunneling conductance is observed below
the lowest bias voltage ($V \approx 1 mV$). This is
 due to the half-metallicity of Co$_2$MnSi. 
However, the tunneling 
conductance rapidly increases in the low-bias region ($10 mV < |V| < 150 mV$),
producing a rapid decrease of the TMR ratio as can be seen in Fig\ref{mtj-Co2MnSi}(b). 
With increasing temperature, the tunneling conductance around the zero-bias region 
gradually increases and the steep structure at low bias  is gradually lost.

\begin{figure}[h]
\includegraphics[width=0.80\linewidth, height=6cm]{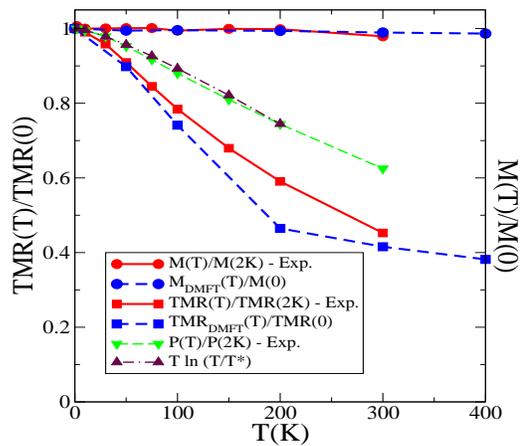}
\caption{(Color online) 
Comparison of the TMR ratio and magnetisation ($M$) as a function of temperature
for a FeCo/Al-O/Co$_2$MnSi tunnel juction as obtained from experimental data 
and from LSDA+DMFT.
Triangle-down/up represents the measured polarization and the fit 
to the analytic expression
$1-P(T)/P(0) \propto T \ln{T/T^*}$ \cite{ir.ka.06}.
}
\label{comp}
\end{figure}
These results are consistent with the 
temperature dependence of the TMR ratios  
and the saturation magnetization for Co$_2$MnSi
plotted in Fig. \ref{comp}.
Here, the theoretical TMR ratio was obtained from the LDA+DMFT polarisation
by using Julliere's formula and assuming 
a temperature-independent 
value of $0.5$ for the spin polarisation of FeCo, as obtained
independently from other measurements~\cite{mo.pa.00}.
Despite its approximate character,  this formula
reproduces correctly the temperature dependence of the spin-polarization
as was previously seen in detailed theoretical investigation of the tunneling
current in the Half-metallic ferromagnets 
\cite{mc.fa.02,ir.ka.06}. 
Moreover, this is the standard formula used to
 extract the spin-polarization from TMR data
\cite{zu.fa.04}. 
The saturation magnetization 
(M$_S$) was measured using a SQUID magnetometer for a MgO-sub./Cr/Co$_2$MnSi film 
having the same bottom electrode structure as the MTJs. 
As one can see,
M$_S$ values change little in the temperature range 
$2 K \le T \le 300 K $ (since $T_c\simeq 985 K$ this corresponds to
$0.002 \le T/T_c \le 0.3 $). In contrast, the spin polarization 
(and thus the TMR ratio) decrease drastically 
in the same temperature range.
As can be seen from Fig.~\ref{comp} the experimental temperature
dependence of the TMR ratio   
and of the magnetisation is
 in rather good agreement with our LSDA+DMFT
calculation. 
Moreover, for $T\lesssim 200K$, the experimental polarisation curve
(obtained from the TMR data by inversion of Julliere's formula)
is quite well reproduced 
by the analytic expression $1-P(T)/P(0) \propto T \ln{T/T^*}$
predicted in Ref.\cite{ir.ka.06}.
Here, $T^*$ represents 
a crossover temperature $\approx (\Delta/W)^2 T_c$,
below which the $\delta P(T) \propto \delta <S^z>$ behavior
is expected.
$\Delta=0.4eV$ is the minority spin gap,
$W \approx 8eV$ the bandwidth and $T_c \approx 985K$.
The value of $T^*$ estimated by this expression ($T^* \approx 2.5 K$)
is in good agreement with the value $T^*=2.7K$
obtained from a fit to the experimental data. 
These facts, 
combined with the 
strong temperature dependence 
of the differential tunneling conductance
shown in Fig.~\ref{mtj-Co2MnSi} clearly
support the existence of minority-spin NQP-states  above the Fermi level.

\begin{figure}[h]
\includegraphics[width=0.80\linewidth]{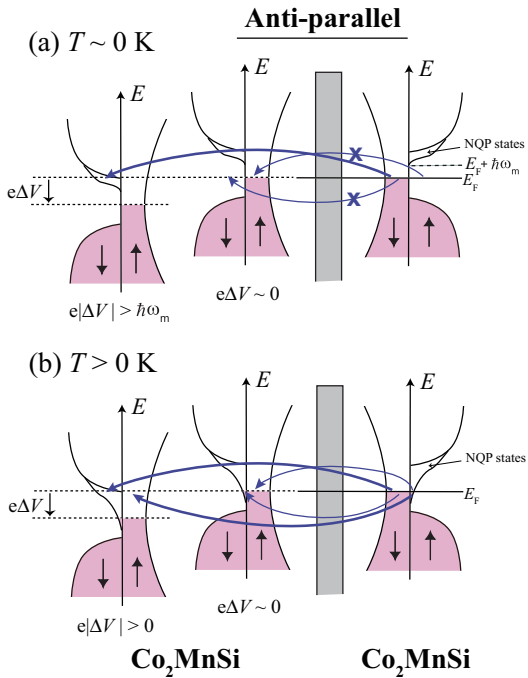}
\caption{(Color online) Schematic representation of an electron tunneling process at
  $ \Delta V\approx 0$ and at finite bias voltage in 
a Co$_2$MnSi/Al-O/Co$_2$MnSi magnetic tunnel junction. At $T\approx
0K$ (a), 
NQP states vanish 
at $E_F$, while for $T>0$ (b) (see text) 
NQP states extends across $E_F$ and an additional tunneling
channel opens.}
\label{AP_scheme}
\end{figure}

The behavior of $dI/dV$ as a function of bias voltage V and
temperature can be understood  
by using the schematic picture shown in Fig. \ref{AP_scheme}. Here,
the crucial role of NQP  
states is apparent. In the case of an ideal half-metal at $T \approx
0K$, no tunneling process  
can occur in the AP state, as there are no electronic states which can contribute to 
the tunneling. Upon applying a finite 
bias voltage ($e |\Delta V| >\hbar \omega_m$ )
to the tunnel junctions,
a conducting channel for the minority spins 
opens  due to the nonvanishing NQP density of states (Fig. \ref{AP_scheme}(a)). 
Here, $\hbar \omega_m$ is the anisotropy gap in the magnon spectrum,
below which the density of non-quasiparticle states  vanishes~\cite{ir.ka.02}. 
At finite temperatures ($k_B \ T > \hbar \omega_m$), 
NQP states are expected to broaden and 
to extend across $E_F$ as shown in Fig.\ref{dos-Co2MnSi}. Therefore, a finite tunneling 
conductance occurs even at vanishing bias  voltage  
(Fig. \ref{AP_scheme}(b)), producing a rapid decrease of  spin polarization. 

The NQP picture suggests that in order to improve the performances of  Co$_2$MnSi-based
MTJ, a special attention should be paid to magnon excitations. 
This suggests a strategy to improve the performances of 
Co$_2$MnSi-based MTJ. The idea is to
modify the magnon excitations,  while at the same time
preserving the electronic properties, i.e. maintaining the gap in the minority channel.
One possibility would be to increase 
the magnetic anisotropy and thus the magnon gap \cite{at.fa.04}. 
This can be achieved by a proper doping with rare-earth 
atoms in the half-metallic material \cite{at.fa.04}. Experimental work is in progress.

Finally, let us comment on alternative depolarisation mechanisms.
Dowben et. al. \cite{sk.do.02} showed that finite-temperature non-collinearity 
produces a spin mixing which ultimately leads to a nonvanishing but
symmetric DOS around the Fermi level in the gap of the insulating spin channel.  
In contrast, as discussed above,
the low-energy DOS within the minority gap induced by
many-body effects (NQP states) appears only above the Fermi level, and
is thus strongly asymmetric. In addition, its low-energy part is 
strongly temperature dependent, and much larger in magnitude than the one produced by 
non-collinear or spin-orbit effects.
Therefore, we expect depolarisation due to NQP states to be dominant in
comparison with other effects, such as as the static 
non-collinearity \cite{sk.do.02} or spin-orbit coupling.


In summary, we caried out  a combined theoretical  
and experimental study of depolarisation 
effects in half-metallic Co$_2$MnSi. Our tunneling conductance measurements in 
 Co$_2$MnSi-based
 magnetic tunnel junctions showed for the first time
the existence of NQP states  above the Fermi level in the minority spin channel. 
The behavior of the finite-temperature conductance demonstrates the important role 
played by NQP states in inducing depolarization, an effect which should be carefully 
considered in designing Co$_2$MnSi-based MTJ devices. The  drastic reduction of the 
TMR ratio from low to room temperatures is a clear evidence of the  detrimental effect
played by the NQP states on the electron spin polarization.
A possible  strategy to improve the performances
of the Co$_2$MnSi-based MTJ
by increasing the magnon anisotropy
 was also discussed.

This study was supported by the IT Program of the RR2002, the CREST
program  of the JST and the NEDO.
L.C. and E.A. acknowledge financial support by the  
FWF P18505-N16,  M.I.K. acknowledges financial support from
FOM (The Netherlands). 
A.I.L. acknowledge financial support from the DFG (Grants No. SFB 668-A3).
Useful discussion with Prof. R.A. de Groot are
appreciated.  

\bibliographystyle{prsty}
\bibliography{references_database}
\end{document}